\documentclass{appolb}
\usepackage{graphicx}
\begin{document}
\title{An event mixing method with invariant-mass/energy hierarchy correspondence cut for Bose-Einstein correlations in $\pi \pi X$ system%
%\thanks{Presented at ...}%
}
\author{Q. He
\thanks{Corresponding author: heqh@nuaa.edu.cn}%
\address{Department of Nuclear Science and Engineering, Nanjing University of Aeronautics and Astronautics (NUAA), Nanjing 210016, China}
\\
%{Third Author of different affiliation
%}
%the Name(s) of other Author(s)
%\address{affiliation}
}
\maketitle
\begin{abstract}
A new event mixing constraint, namely invariant-mass/energy hierarchy correspondence (IMEHC) cut, is introduced for the low-multiplicity event mixing technique for the purpose of measuring Bose-Einstein correlations (BEC) in exclusive reactions with $\pi \pi X$ final state particles. The mixing cut is relevant to the hierarchy of the invariant mass of $\pi X$ system and two bosons’ energy hierarchy. Numerical tests are performed to check the validity of the new mixing method. As long as the measurements of BEC parameters $r_0$ and $\lambda_2$ are considered, this new mixing method is effective to observe BEC effects and the systematic bias of $r_0$ and $\lambda_2$ are smaller than the previously proposed mixing cut.
\end{abstract}

  
\section{Introduction}
Bose-Einstein correlations (BEC) \cite{Weiner1999,Boal1990}, generally applied in high-energy elementary particle collisions and relativistic heavy ion collisions for measuring space-time properties of particle production volume \cite{Csorgo2006,Utyuzh2007,Ma2007,Xu2017, Liu2012, Brown2007, Heister2004, Khachatryan2010}, can also be used in exclusive reactions with low multiplicity to measure the spatial extension of excited baryons generally decayed back to the ground states with two-meson emission, such as the reaction $\gamma p\to N^*\to \pi^0 \pi^0 p$ at incident photon energies around 1 GeV. However, such studies are still unavailable, because the event mixing method \cite{Kopylov1972_15, Kopylov1974} used for BEC measurement for low multiplicity reactions is strongly disturbed by global conservation laws and resonance decays  \cite{Klaja2010, Chaje2008} which may lead to significant non-BEC kinematical correlations of final state particles and make the BEC observation more complicated. 

It is still a challenging work to find appropriate event mixing method to suppress the influence of kinematical correlations arising from global conservation laws and intermediate resonances. In 2016, a mixing technique with two mixing cuts, named missing mass consistency (MMC) cut and pion energy (PE) cut, is proposed for $\pi^0 \pi^0$ BEC measurement in the $\gamma p\to \pi^0 \pi^0 p$ reaction at incident photon energies $E_{\gamma}$ around 1 GeV (a non-perturbative QCD region) \cite{He2016}. The MMC cut that requires the missing mass in the mixed event should be identical to that of the original event, was introduced for the sake of energy momentum conservation in the mixed events. The PE cut that rejects some events with boson energy beyond a certain level, is used to avoid two-pion energy sum exceeding physically allowed limits. Because the PE cut results in sample reduction (about 40\% proportion) and hence leads to a worse analysis accuracy, a new cut, named the energy sum order (ESO) cut  \cite{He2018} was proposed later to replace the PE cut, which has no requirement on discarding original events. However, as long as the fit BEC parameters are considered, the ESO cut has big systematic errors for BEC parameter measurement.

In this work, a new mixing constraint, named invariant-mass/energy hierarchy correspondence (IMEHC) cut is introduced to improve the accuracy of BEC parameter measurement. The IMEHC cut employs the hierarchy of the invariant mass of $\pi X$ system and two bosons’ energy hierarchy to control the mixing process. Extensive numerical tests are carried out to test the ability of the new mixing cut IMEHC to observe BEC effects.

\section{Event mixing method with invariant-mass/energy hierarchy correspondence cut}
To employ BEC effects to investigate the space-time properties of subatomic reacting volume emitting identical bosons, one needs to measure a two-particle correlation function  \cite{Alexander2003,Goldhaber1960}
\begin{equation}
C_{BEC} (p_1,p_2 )=\frac{P_{BEC} (p_1,p_2)}{P_0 (p_1,p_2)}=1+|f(q)|^2,                                         
\label{eqn1}
\end{equation}
where $P_{BEC} (p_1,p_2)$ stands for the probability of emitting two identical bosons with momenta $p_1$ and $p_2$ with BEC effects, and $P_0 (p_1,p_2 )$ the probability of so-called “reference sample” without BEC effects. $f(q)$ is the Fourier transform of the emitter source distribution, where $q=p_1-p_2$ ( $p_1$ and $p_2$ are two bosons’ momenta). If the emitter source has a Gaussian density distribution, Eq. (\ref{eqn1}) is parametrized in terms of a “source radius” $r_0$ and a “chaoticity parameter” $\lambda_2$:
\begin{equation}
  C_{BEC} (p_1,p_2 )=C_{BEC} (Q)=N(1+\lambda_2 e^{-r_0^2 Q^2}),                                                                    
\label{eqn2}
\end{equation}
where $Q$ is a measurement of the relative momentum between two bosons defined by $Q^2=-q^2=-(p_1-p_2)^2$, which is a Lorentz invariant parameter widely used in many of the two-boson one-dimensional BEC analyses. The parameter $N$ is the normalization factor. In this parametrization, the parameter $r_0$ is generally similar ($r_0 \sim$ 0.5-1 fm) for all hadronic interactions (excluding heavy ion interactions), while the parameter $\lambda_2$, a measurement of the boson-emitting chaoticity, varies from 0 to 1 depending upon the method of fit and experimental factors in measuring data sample such as particle misidentification and detecting resolution \cite{Andersson1998}.

    The reference sample free of BEC effects is generally produced from the original data sample through the event mixing technique  \cite{Kopylov1972_15, Kopylov1974}, which eliminates the BEC effects via selecting two bosons’ momenta from two random events in the original data sample under prescribed cut conditions. But the application of event mixing method to exclusive reactions with only two identical bosons is still challenging because it is strongly interfered by non-BEC factors such as global conservation laws and resonance decays. In this case, in order to make a valid reference sample identical to the real data in all aspects but free of BEC effects, special mixing constraints are required in event mixing. With an ideal event mixing method a reference sample should have identical $Q$ distribution to the original one and hence obtain a flat correlation function.
 
  The knowledge of kinematical correlations between final state particles in original samples may provide useful information for appropriate constraints to govern event mixing process. Inspired by the idea in Ref. \cite{He2016} that one pion with relatively higher/lower energy can only be swapped with another pion from another event with relatively higher/lower energy in order to maintain the original kinematical correlations of two pions in the sequential decay reactions $\gamma p \to \pi^0 \Delta \to \pi^0 \pi^0 p$, in this work a new mixing constraint, named invariant-mass/energy hierarchy correspondence (IMEHC) cut, is proposed for measuring BEC effects in exclusive reactions with $\pi \pi X$ final states.

The IMEHC constraint contains two sub-cuts. The first sub-cut is relevant to the invariant mass of $\pi X$ system among the final state $\pi \pi X$, defined by $m^2 (\pi,X)=(p_\pi+p_X)^2$. It requires to exchange two pions both with lower/higher invariant mass $m^2 (\pi,X)$ from two different events. The second sub-cut governs the mixing procedure in terms of the energy of pion. It requires one pion with relatively higher/lower energy can only be swapped with another pion from another event with relatively higher/lower energy. In mixing procedure at a time, only one sub-cut of the IMEHC constraint is randomly selected with equal probability to govern the event mixing, while another is temporarily in active. In other words, the two bosons being swapped should be equal in invariant mass $m(\pi,X)$ hierarchy at a time or in energy hierarchy at another time. In the new mixing method, the MMC cut \cite{He2016} and the energy sum order (ESO) cut \cite{He2018} are still included in the mixing method. The ESO cut is expressed by the definition $min⁡(E_{sum}^{(ori,1)},E_{sum}^{(ori,2)} )<E_{sum}^{mix}<max⁡(E_{sum}^{(ori,1)},E_{sum}^{(ori,2)})$   \cite{He2018}, where $E_{sum}^{(ori,1)}$ and $E_{sum}^{(ori,2)}$ are the two-boson energy sums in the two original events, $E_{sum}^{mix}$ the same value in the mixed event.

\section{Numerical verification}

Numerical simulation is performed to test the mixing cut IMEHC for measuring two-pion Bose-Einstein correlations in reactions with $\pi \pi X$ final state particles. In the simulation, the reaction $\gamma p\to \pi^0 \pi^0 p$ with and without BEC effects is taken as an example to demonstrate the event mixing method.  The event samples are generated using a ROOT \cite{Brun1997} utility named “TGenPhaseSpace” \cite{James1968}. The details of the event generation can be found elsewhere \cite{He2016}. Totally six $\gamma p\to \pi^0 \pi^0 p$ event samples with and without BEC effects at incident photon energies of 1.0 GeV, 1.03 GeV, 1.06 GeV, 1.09 GeV, 1.12 GeV, and 1.15 GeV respectively are generated.
  
\begin{figure}[htb]
\centerline{%
\includegraphics[width=0.57\linewidth]{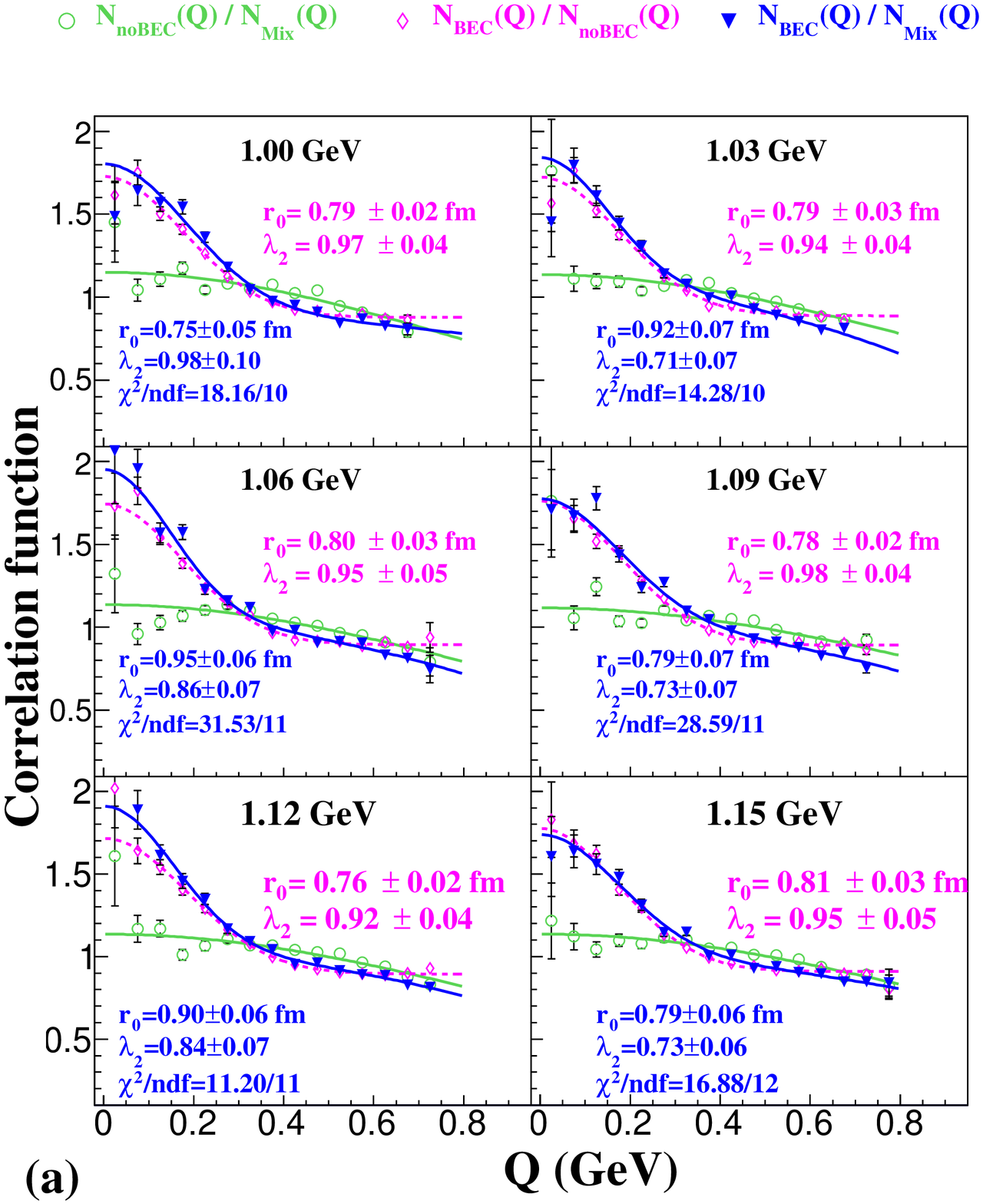}
\includegraphics[width=0.43\linewidth]{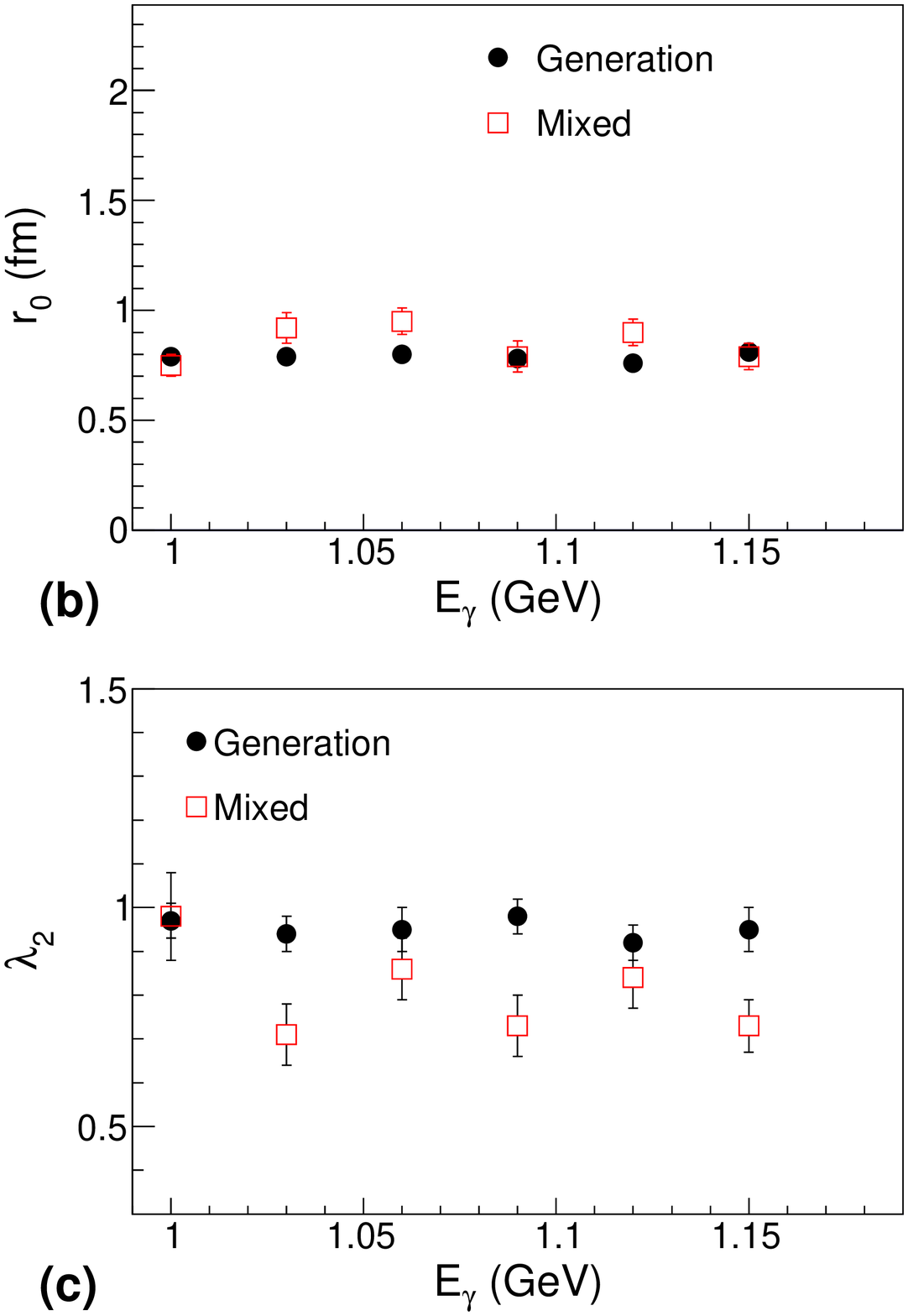}
}
\caption{(a) Correlation functions of the mixed events ($N_{BEC} (Q)/N_{Mix} (Q)$). For comparison, the correlation functions of the generated BEC samples ($N_{BEC} (Q)/N_{noBEC} (Q)$) and those of the non-BEC samples ($N_{noBEC} (Q)/N_{Mix} (Q)$) are also presented. (b) Fitted BEC parameters of $r_0$ obtained by the proposed event mixing method at six incident photon energies $E_{\gamma}$=1.0, 1.03, 1.06, 1.09, 1.12, and 1.15 GeV for the $\gamma p \to \pi^0 \pi^0 p$ events. (c) Fitted BEC parameters of  $\lambda_2$. For comparison, the values of $r_0$ and $\lambda_2$ for the generated sample with BEC effects are also shown. }
\label{Fig1}
\end{figure}

  Although the obtained correlation functions of the non-BEC samples have not any enhancements at $Q=0$ and have a semi-flat distribution averagely as shown in Fig. \ref{Fig1} (a), they exhibit a $Q^2$ dependent pattern. Thus the quadratic function $f(Q)=N(1+αQ^2)$ is used to fit the non-BEC correlation functions. Because of the $Q^2$ dependent pattern of the non-BEC sample correlation functions has a strong association with the BEC-samples correlation function fitting, the later should be fitted by a modified Eq. (\ref{eqn2}):
\begin{equation}
 C_{BEC} (Q)=N(1+\alpha Q^2)(1+\lambda_2 e^{-r_0^2 Q^2}),                                                                    
\label{eqn3}
\end{equation}
                                                                                              
The ability of the proposed mixing method to measure BEC effects is tested using the six BEC samples with input BEC parameters typically set to be $r_0=0.8$ fm and $\lambda_2=$1.0. As shown in Fig. \ref{Fig1} (a), the BEC effects can be successfully observed in the obtained correlation functions using the proposed mixing method. 

The BEC parameters $r_0$ and $\lambda_2$ determined from the proposed mixing method are found to be consistent with the input values of the generated BEC samples within error bars at most energy bins, as shown in Fig. \ref{Fig1} (b) and (c). Because of the event mixing induced $Q^2$ dependent pattern of the non-BEC sample correlation functions, the BEC parameters from event mixing are determined by fitting Eq. (\ref{eqn3}) to the event mixing obtained correlation function ($N_{BEC} (Q)/N_{Mix} (Q)$). As not involved in event mixing, the input BEC parameters used for comparison are obtained by fitting Eq. (\ref{eqn2}) to the correlation functions of the generated BEC samples ($N_{BEC}(Q)/N_{noBEC} (Q)$). No dependence on the incident photon energy is found for both  $r_0$ and $\lambda_2$. Summing over the six energy bins, the error weighted mean value of  $r_0$ is determined to be $0.84\pm0.03$, about 8\% overestimated compared to the mean value of the input one, $0.78\pm0.01$, and that of $\lambda_2$ is found to be $0.78\pm0.03$, about 18\% underestimated compared to the mean value of the input one, $0.95\pm0.02$. 

\begin{figure}[htb]
\centerline{%
\includegraphics[width=0.5\linewidth]{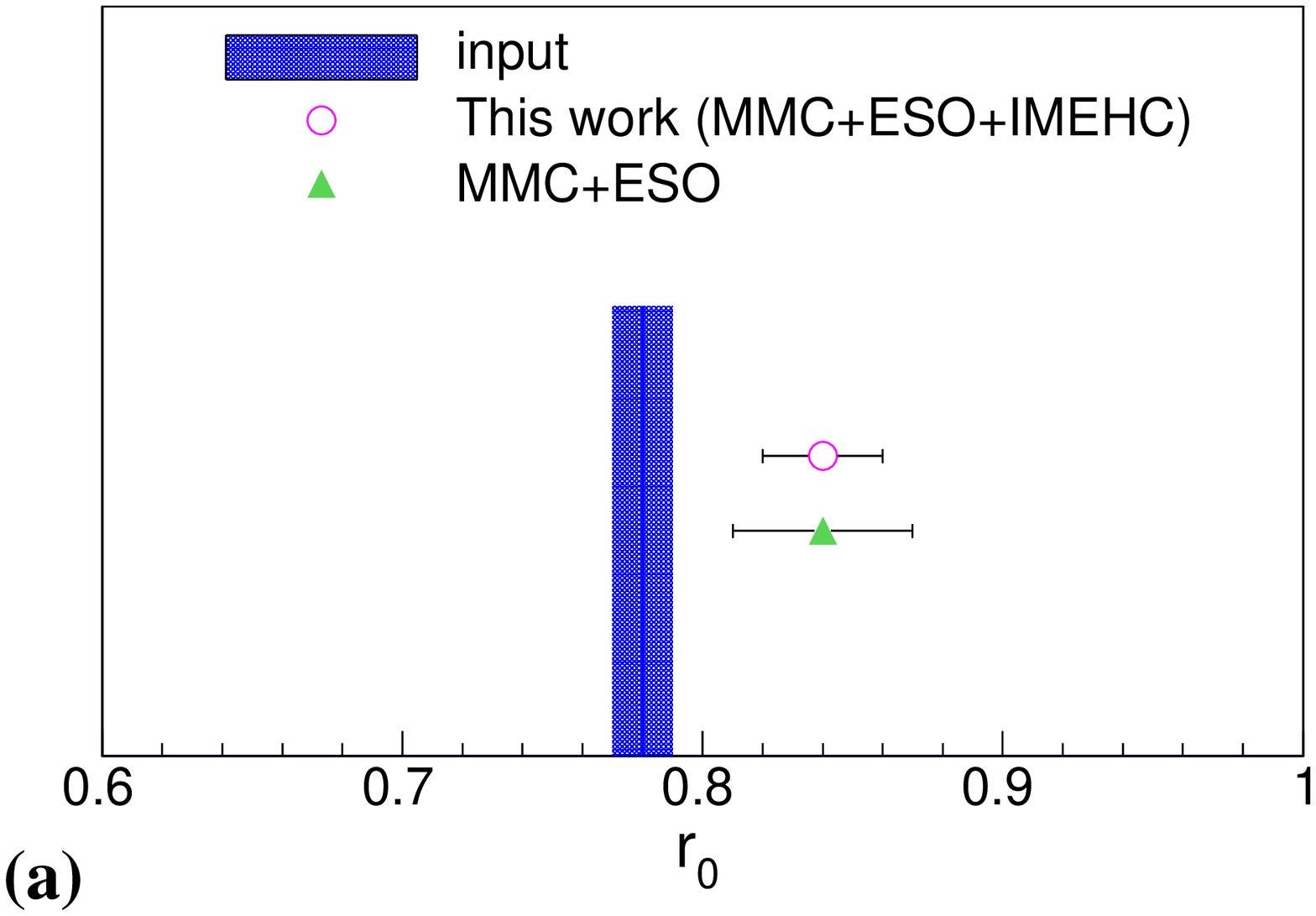}
\includegraphics[width=0.5\linewidth]{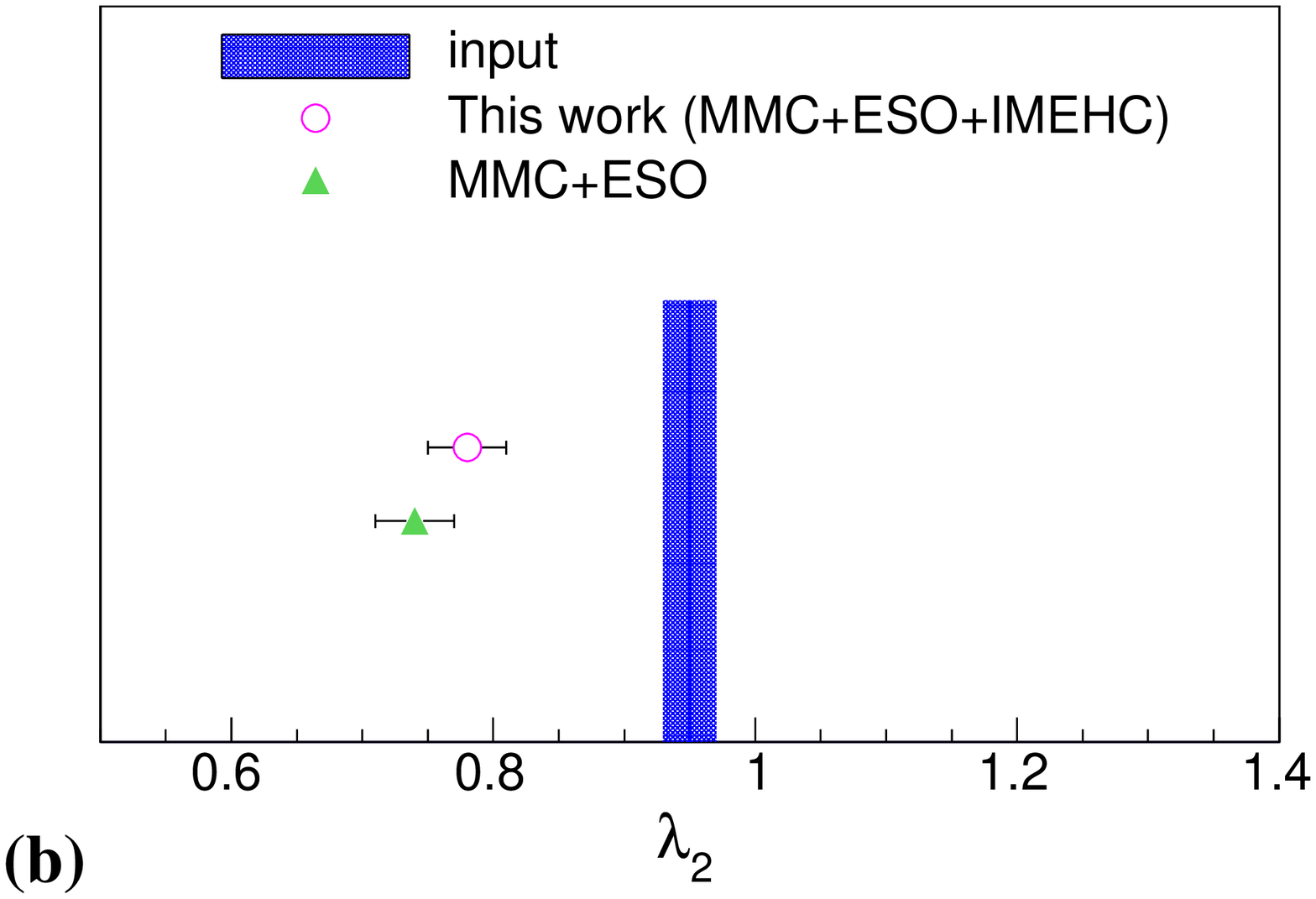}
}
\caption{Comparing input BEC parameters with those from the proposed mixing method. For comparison, the results from the mixing method using the MMC and ESO cuts \cite{He2018} are also presented. }
\label{Fig2}
\end{figure}

By comparing the mean values of the BEC parameters obtained with the mixing method with those from the previously proposed mixing method using only the MMC and ESO cuts \cite{He2018}, it is found that the $r_0$ remains the same as the previously proposed method, while the mean value of $\lambda_2$ is closer to the input one. Compared to the previously proposed mixing method, this new mixing method reduces the systematic bias of $\lambda_2$ from 22\% to 18\% as shown in Fig. \ref{Fig2}. As for the uncertainties of the BEC parameters, $r_0$ has a smaller uncertainty compared to that from the previously proposed mixing method, while the $\lambda_2$ uncertainty remains the same. From this point, the improvement of this new method is still very limited, and further improvement is needed in the future studies.

Although the IMEHC cut improves the systematic bias for BEC parameters fitting, this method still introduces systematic bias somehow. The fit BEC parameters $r_0$ and $\lambda_2$ obtained by the proposed mixing method should also be corrected in practical application. In addition, the $Q^2$ dependent fitting problem still remains. Therefore future efforts should concentrate on improving the accuracy and searching new mixing cuts to bypass the $Q^2$ dependent fitting procedure. 
  
\section{Summary}
Because general event mixing techniques developed for inclusive reactions at high energies with a large multiplicity cannot be directly applied to exclusive reactions at low energies with a very limited multiplicity, in this work an event mixing technique equipped with a new constraint, named invariant-mass/energy hierarchy correspondence (IMEHC) cut, is proposed especially for Bose-Einstein correlations (BEC) effects measurements in reactions with $\pi \pi X$ final state particles. The accuracy of BEC parameters observation with this new mixing cut is verified using numerical simulations with $\gamma p\to \pi^0 \pi^0 p$ events with and without BEC effects. It is found that the extracted mean value of $r_0$ is about 8\% overestimated and $\lambda_2$ is averagely about 18\% underestimated for typical input BEC parameters $r_0$=0.8 fm and $\lambda_2=$ 1.0. The systematic bias of $\lambda_2$ is smaller than the previously proposed mixing cut using only the MMC and ESO cuts \cite{He2018} (reduced from 22\% to 18\%). Future efforts are still needed to improve the accuracy and to find new mixing cuts aiming to avoid the $Q^2$ dependent fitting procedure in order to get rid of additional fitting parameter to fit the $Q^2$ dependent pattern.

\section{Acknowledgements}
This work was supported by the National Natural Science Foundation of China (No. 11805099) and the Fundamental Research Funds for the Central Universities (Nos. 1006-XAA18059, 1006-YAH17063).

%uncomment the following lines to place a figure
%\begin{figure}[htb]
%\centerline{%
%\includegraphics[width=12.5cm]{Fig1}}
%\caption{Plot of ...}
%\label{Fig:F2H}
%\end{figure}

%\bibliographystyle{elsarticle-num}
%\bibliography{ref}

\end{document}